# Changes of statistical structural fluctuations unveils an early compacted degraded stage of PNS myelin

Nicola Poccia[1,2,*], Gaetano Campi[3,*], Alessandro Ricci[4,1], Alessandra S. Caporale[1], Emanuela Di Cola[5], Thomas A. Hawkins[6], Antonio Bianconi[1,3]

[1]*RICMASS Rome International Center for Materials Science Superstripes, via dei Sabelli 119A, 00185 Roma, Italy.*
[2]*MESA+ Institute for Nanotechnology, University of Twente, P. O. Box 217, 7500AE Enschede, The Netherlands.*
[3]*Institute of Crystallography, CNR, via Salaria Km 29.300, 00015 Monterotondo Roma, Italy.*
[4]*Deutsches Elektronen-Synchrotron DESY, Notkestraße 85, D-22607 Hamburg, Germany.*
[5]*European Synchrotron Radiation Facility, B. P. 220, F-38043 Grenoble Cedex, France,*
[6]*Department of Cell and Developmental Biology, UCL, Gower Street, WC1E 6BT London, UK.*

**Abstract**   Degradation of the myelin sheath is a common pathology underlying demyelinating neurological diseases from Multiple Sclerosis to Leukodistrophies. Although large malformations of myelin ultrastructure in the advanced stages of Wallerian degradation is known, its subtle structural variations at early stages of demyelination remains poorly characterized. This is partly due to the lack of suitable and non-invasive experimental probes possessing sufficient resolution to detect the degradation. Here we report the feasibility of the application of an innovative non-invasive local structure experimental approach for imaging the changes of statistical structural fluctuations in the first stage of myelin degeneration. Scanning micro X-ray diffraction, using advances in synchrotron x-ray beam focusing, fast data collection, paired with spatial statistical analysis, has been used to unveil temporal changes in the myelin structure of dissected nerves following extraction of the Xenopus laevis sciatic nerve. The early myelin degeneration is a specific ordered compacted phase preceding the swollen myelin phase of Wallerian degradation. Our demonstration of the feasibility of the statistical analysis of SµXRD measurements using biological tissue paves the way for further structural investigations of degradation and death of neurons and other cells and tissues in diverse pathological states where nanoscale structural changes may be uncovered.

**Introduction**

A full and complete understanding of peripheral nerve degeneration is essential to help advance of the clinical management of sharp nerve injuries [1], peripheral neuropathic conditions such as Charcot Marie Tooth disease [2] and in relation to axonal and neuronal damage in demyelinating activity in Multiple-Sclerosis lesions [3,4]. The peripheral nervous system (PNS) controls the transmission of information via nerve signals from the central nervous system to the periphery and vice-versa. The PNS consists of a complex network of peripheral nerves containing myelinated and unmyelinated axons. The degeneration of PNS axons distal to a lesion is a complex process involving multiple molecular and cellular events. The process was first described in the frog by Augustus Waller [5] and the well documented sequence of events now bears his name [6-8]. The Wallerian degeneration of PNS nerves [6-9], can first be observed by the degeneration of axons 24-48 hours after a lesion, from that point, the completion of the process takes 3 – 6 weeks. The disintegration of axons is usually completed within 7–10 days. This has been assigned to activation of axoplasmatic proteolyses, which occurs as a response to intracellular calcium influx [7] and to the activation of ubiquitin-proteasome system [8].

To provide insight into the early process of degeneration of myelinated axons within the first day after an injury, before Wallerian degeneration is normally observable, the development of an increased level of spatial structural resolution is needed, using non-invasive methods to bear in the biological realm. We have focussed our studies on this early stage of PNS degeneration by examining the myelin sheath of axons in the sciatic nerve of Xenopus frogs, a tissue that has been used previously to investigate the PNS degeneration [10-11].

Myelin is an elaborate multilamellar membrane surrounding many of the axons of the PNS it is fundamental for normal nervous system function [12-16]. Structural studies of myelin often use the sciatic nerve as representative of the PNS. Many decades of research into the structure-function relationship of myelin have led to the understanding that saltatory conduction by a myelinated axon is mediated by decreased capacitance and increased electrical resistance of the internodal axonal membrane, it is also enhanced by the clustering of sodium channels at the nodes of Ranvier [13]. Myelin ultrastructure has been studied using a wide variety of approaches such as transmission electron microscopy (TEM) [14], coherent Anti-Stokes Raman scattering microscopy [15] and neutron scattering [16,17]. Small angle X ray diffraction (XRD) is widely used for structural studies of biological tissues and processes [18] and it is well suited to study myelin because of the quasi-periodic structure of myelinated axons [19-21]. For example myelin sheath structural modifications caused by mutation of P0 glycoprotein, have been revealed using XRD by Avila *et al* [21].

Myelin is, from a morphological point of view, a very heterogeneous tissue with intrinsic fluctuations of the structural order from point-to-point in real space: this makes quite difficult to probe its intrinsic structural fluctuations with conventional experimental approaches. For example, XRD is limited since it provides insight only into the periodic molecular structure of myelin probing the k-space (or reciprocal space) with no spatial resolution. Conventional transmission electron microscopy (TEM) is a local, highly spatially resolved approach, but it suffers from sample fixation and dehydration artifacts making it impossible to use in the determination of functional spatial structure fluctuations. Therefore we need to develop non-invasive probes with high spatial structure resolution.

Recently advances in synchrotron beam focussing have allowed the development of new non-invasive special probes: micro x-ray diffraction (*μ*XRD) [23-28] and micro X-ray Absorption Near Edge Structure (XANES) [29]. These microscopies have already been used to provide unique information in different research fields ranging from imaging fibers [23], monitoring cerebral myelin [24-26], bone tissue mineralization [26,27] and bone quality in implants [28]. Up to now, mapping of the variations of the structure of myelin sheaths in the brain has been achieved [24,25] with a spatial resolution of 50x50 μm$^2$ by De Felici et al. [24]. They probed white matter myelin in a total of 20 different locations in differently aged subjects. Jensen et al. [25] mapped the concentration and periodicity of myelin sheaths in a rat brain by tomography, looking for correlations between changes in the tracts of the brain and neurodegenerative diseases. Ducic et al. have also studied hereditary neuropathic effects on the structure of myelin [26].

Here we report the application of a novel approach to the examination of peripheral nerve myelin and its early degradation by focusing on the determination of "statistical distribution of local lattice fluctuations" using fast scanning SµXRD. This method uses SµXRD to probe both the k-space (reciprocal space) and the real space inhomogeneity of complex materials. The statistical distribution of the fluctuating structural order is extracted by methods of statistical physics applied to data analysis. This approach has been recently used to get key information on the relation between material functionality and the "statistical distribution of multi-scale structural fluctuations" from the nanoscale to micron scale due to defects in complex functional materials [30-35]. These studies have related multi-scale, scale free structural fluctuations [30,33,35] and their time evolution [31] with material functionality. Scale free distribution has been recently shown to play a key role in cell biology and brain functionality [36-40].

We first determine the statistical distribution of the intrinsic spatial heterogeneity of myelin structure fluctuating from point to point in a freshly extracted unfixed nerve. We then examine variation in these metrics in a longer-term cultured unfixed nerve, looking for any changes potentially caused by degradation in the nerve in the first 24 hours following extraction, before the onset of early Wallerian degeneration.

We collected the spatial fluctuations of the structure of myelin lamellae at thousands of discrete locations using a short time scale for data collection. The short time collection data of SµXRD microscopy was methodologically possible by using bright synchrotron radiation, active X-ray focusing optics and the use of a precise micro-goniometer.

To the best of our knowledge, this is the first work that provides the statistical spatial fluctuations of myelin nanoscale structural order by applications of statistical physics methods. This is a significant methodological advance in its own right. Additional to this we have obtained highly suggestive evidence (albeit with a minimal sample size) that in the early stages of degradation the average myelin multilamellar period decreases.

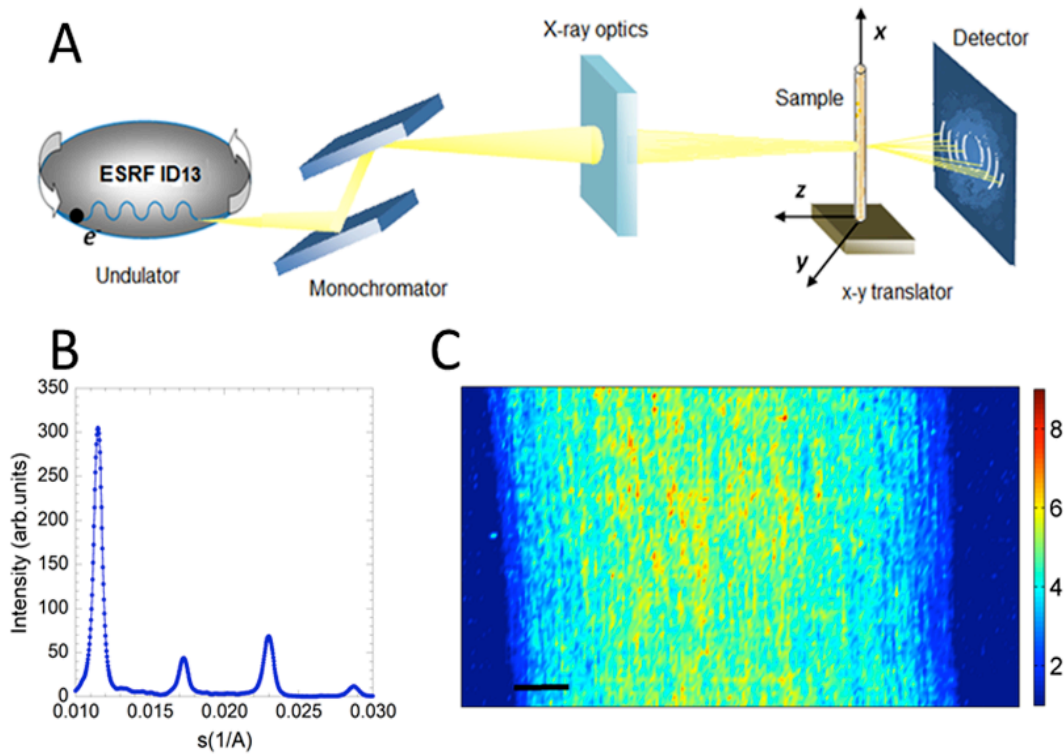

**Figure 1: (A)** The experimental apparatus is located at the ID13 beamline of the European Synchrotron Radiation Facility (ESRF) and features an electron undulator providing 12.4 keV X-rays to crystal optics delivering a 1 µm beam spot on the sample. The sample holder houses a 1 mm-diameter quartz capillary containing the nerve and the culture medium. The x-y translator allows the sample to move in both horizontal and vertical directions with a sampling step of 5 µm within a selected area. The X-rays scattered by the sample are recorded by a FReLoN detector. **(B)** Intensity of the x-ray diffraction reflections after background subtraction, with respect to the impinging beam of the sciatic nerve bathed in normal Ringer's solution at pH 7.3 for the native nerve. The intensity (in arbitrary units) is plotted against the reciprocal distance (Å$^{-1}$). **(C)** Colormap of the integrated intensity of the myelin XRD diffraction pattern measured point by point of a full native nerve. The intensity is greater in the centre and decreased at the edges of the nerve due to the differing quantities of myelinated axons traversed by the beam in these different locations. This is caused by the generally cylindrical shape of the nerve. The bar corresponds to 100 µm.

This myelin compaction process is accompanied by an increase in diffraction intensity. Therefore, the data reveal the formation of a more compacted and ordered, rigid phase showing that a first stage of degradation occurs before the onset of the Wallerian swelling and disordering which is classically described as starting only after 24 hours.

**Results and Discussion.**

The experimental apparatus for SµXRD is shown schematically in **Figure 1A**. The synctrotron radiation emitted by an undulator at the European Synchrotron Radiation

Facility (ESRF) is first monochromatized to get a photon beam of 12.4 keV which is focused on 1x1 µm area. The diffraction pattern at the sample spot illuminated by the x-ray beam is collected in transmission [37, 38]. The nerve is mounted on a y-z translation stages which allows to illuminate different selected spots in the sample. In each 1µm x 1µm measured pixel the ordered multilayer structure of the myelin sheath produces a typical diffraction pattern shown in **Figure 1B**. It was of fundamental importance to minimize any possibility of damage of the nerve. Depending on the radiation dose, cells may be undamaged, damaged and repaired by cell-intrinsic mechanisms, damaged and operating abnormally or they can be killed as a result of the dose. The data acquisition is thus preceded by an estimation of radiation damage to test the actual tissue damage due to the X-ray impinging beam that has a fixed energy. This leads to an estimate of the most suitable exposition time balancing the best signal with the least alteration of the sample. The diffraction pattern was abnormal from an exposure time of 0.9 s, with a broadening of the peaks, and a noisier profile. We decided to use an exposure time of 300 ms did not sacrifice sensitivity, minimizing the radiation damage at the same time. We used scanning step of 4 µm to avoid overlapping effects with the previous illuminated 1µm x 1µm spot.

Through the use of FIT2D (Andy Hammersley, ESRF, Grenoble, France), the diffraction intensity within the rings (Fig. 1A) are radially integrated, and the 1-D X-ray patterns are easily extrapolated. Bragg diffraction orders $h$ = 2, 3, 4, 5 (see Fig. 1B) correspond to the concentric rings, produced by the ordered structure of the myelin multilayer. The XRD diffraction patterns were normalized with respect to the intensity of the impinging beam $I_0$. Background scattering from the empty capillary was measured and subtracted.

A typical full image of the nerve was measured in 20302 pixels (corresponding to a surface of 1 x 0.5 mm$^2$) with a total acquisition time of the scanning procedure of approximately 1 h 40 minutes. In **Fig. 1C** we show the integrated intensity of the X ray diffraction patterns (Fig 1B), recorded at each pixel, after background subtraction and normalization for the incident flux. This intensity is the order parameter for the myelin multilamellar ultrastructure. In the central part of the nerve we observe a stronger intensity, from yellow-to red, while it decreases towards the edges of the nerve. The blue area on the left and right side of the image corresponds to the region where the X-ray beam crossed the edge of the nerve.

Our investigation through SµXRD imaging was aimed to quantify the early changes in the ultrastructure of myelin in the degradation of a native nervous tissue. Specific regions of interest (ROI), were taken from the central part of the nerve where it is thickest and most parallel to the capillary wall in order avoid artifacts due to the smaller numbers of axons near the nerve edge. Typical maps reporting the integral intensity of X ray diffraction spectra in selected 40x120 µm central zones of both *native* and *early degraded* nerve are shown in **Fig. 2A** and **Fig. 2B** respectively.

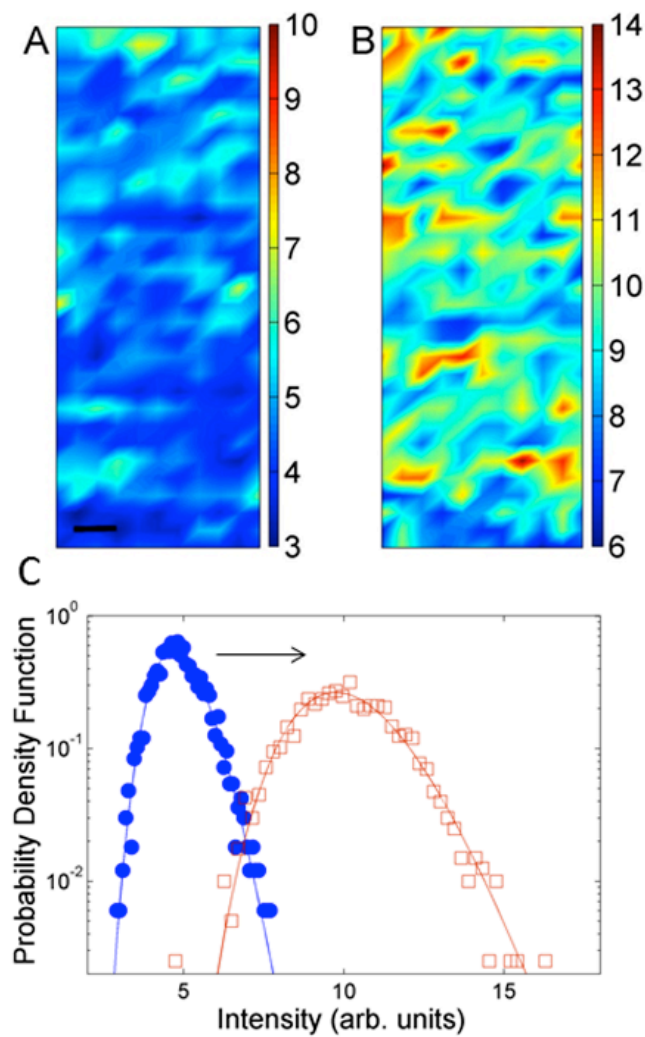

**Figure 2:** Integrated intensity of X ray diffraction patterns measured pixel by pixel in a typical selected (40 x 120 µm) central area, of **(A)** native and **(B)** early degraded nerve. The bar corresponds to 10 µm. **(C)** Probability density function of Integrated intensity measured in the areas (A) and (B) for (full circles) native and (empty squares) early degraded nerve. Both distributions are fit by Lognormal lineshape (solid lines). The average intensity increases by a factor 2 in the early degraded nerve.

The intensity of the XRD signal probes the ordering of the myelin lamellae; therefore this figure gives a spatial map of the reciprocal space i.e. the Fourier transform of the periodic stacking of myelin membrane. At each illuminated spot in the central area of the nerve the signal is the result of the average diffraction of approximately 50 axons which is ideal to get a local averaged statistical value for statistical physics at each spot. The probability density functions of the signals recorded at each illuminated spot in the native and early degraded nerve are shown in **Fig. 2C**. The distributions have a Lognormal line-shape, as shown by the fit. The lognormal distribution is typical of complex systems, [36-39] when the fluctuations of the order of magnitude of the measured physical quantity dominate. This indicates that the fluctuations of the myelin order parameter is characterized by intrinsic spatial correlations in living matter which induce deviations from the random behavior common of glassy materials.

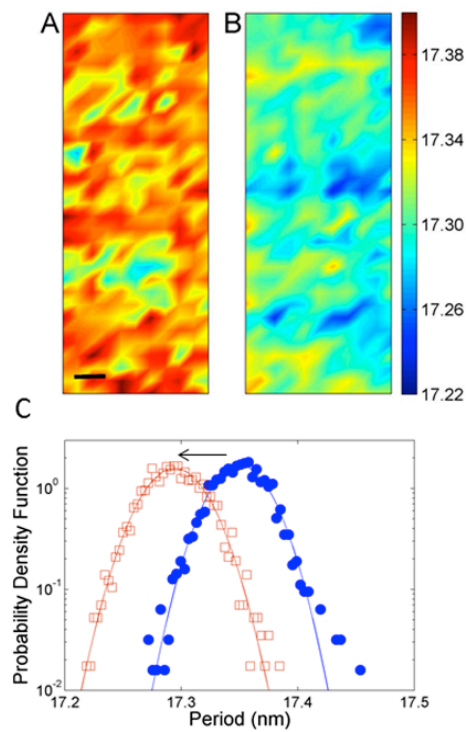

**Figure 3:** Period measured pixel by pixel in a central 40 x 120 µm area, of the **(A)** native and **(B)** early degraded nerve. The bar corresponds to 10 µm. **(C)** Probability density function of period measured in areas (A) and (B) for (full circles) native and (empty squares) early degraded nerve. The periodicity presents intrinsic fluctuations of about 2 Angstroms. We observe a distribution shift of 0.8 ± 0.1 Angstrom, indicated by the arrow, towards lower values for the early degraded nerve.

Going from the native to early degraded state there is a shift in the average integrated intensity towards higher values in the early degraded nerve. This increased average value suggests the occurrence of a crystallization-like process in the myelin membranes.

The period was calculated using standard methods [20,21]; it represents the radial repeating units of the myelin sheath: the lamellae. The radially packed layers include two lipid polar bi-layers that belong to the cellular membrane of the myelin generating Schwann cell, the cytoplasm between these bilayers and the extracellular space included in the sheath during the process of myelination [41]. The average myelin periodicity of the native nerve is found to be 17.36 ± 0.03 nm, in agreement with previous works near physiological ionic strength and at neutral pH [19].

The respective period measured point by point in the ROI for the native and the early degraded nerve is shown in the maps of **Fig. 3A** and **Fig. 3B**. Although both the maps show a consistent inhomogeneity, the probability density function shown in **Fig. 3C** shows a clear shift of 0.8 ± 0.1 Ångström in the average period towards lower values in the early degraded state. This indicates a compaction of the multilamellar structure. The probability density function of the native state shows a fat tail indicating large fluctuations away from the average, probably due to intrinsic structural dynamics.

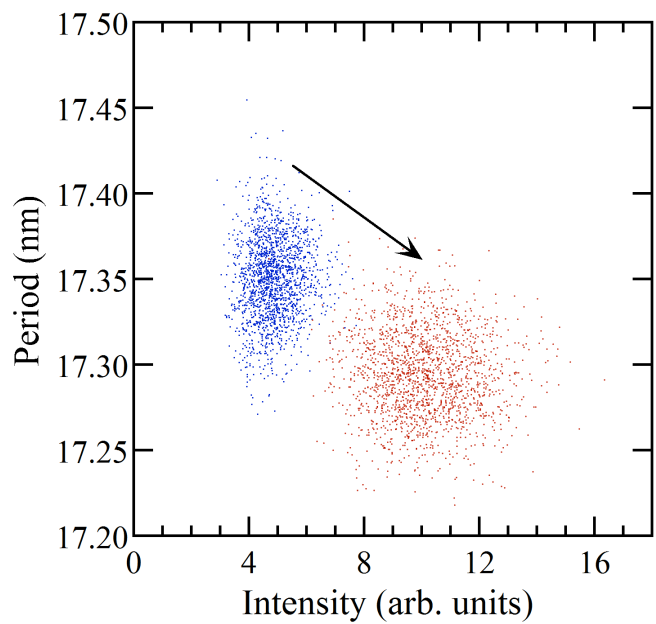

Figure **4**: Scatter plot of myelin period as a function of diffraction intensity for (blue dots) native and (red dots) early degraded nerve.

The correlation between the multilamellar ordering and the myelin period occurring in the system during the early degradation has been characterized through i) the scatter plot of the intensity as a function of the period shown in **Fig. 4** and ii) the two point spatial correlation functions, G(r), in **Fig. 5**. Fig. 4 shows that the decreasing period of the multilamellar structure is correlated with increasing diffraction intensity, typical of compaction towards a more rigid phase.

The G(r) for the native and early degraded nerve are calculated and fitted with an exponential distribution. The correlation lengths of the intensity are 173 µm and 322 µm for the native and the early degraded nerve (upper panel of Fig. 5) respectively. This two-fold increase quantifies the transition to a rigid phase.

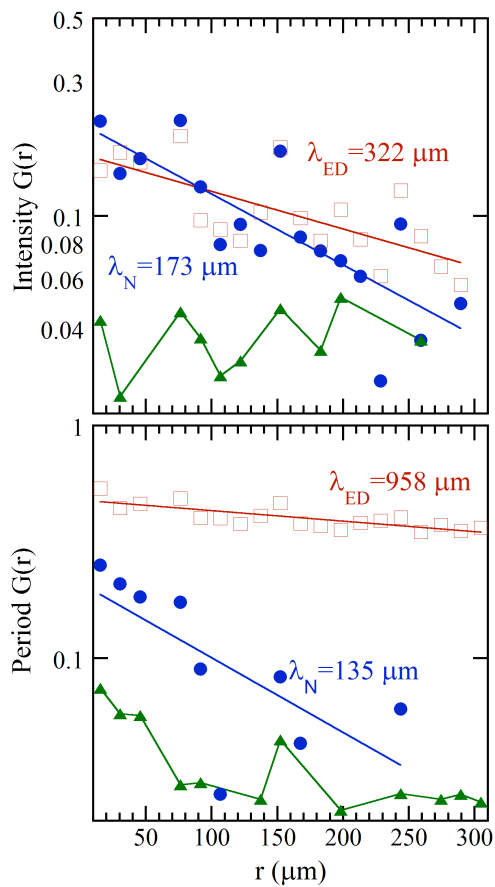

**Figure 5:** Two-point spatial correlation function G(r) for (upper panel) the integrated intensity and (lower panel) myelin period. In each panel the exponential behavior of (full circles) native and (empty squares) early degraded nerve, is reported. The noise level, representing the spatial correlation of the randomized matrices, is reported by triangles.

The correlation length for the myelin period fluctuations in the early degraded nerve is 958 micron, almost ten times higher than the correlation length of 135 micron for the native case. This increase demonstrates the increased regularity in the structural fluctuations over several hundred micron scale in the early degraded respect the native nerve.

**Conclusions.**

We have demonstrated that it is feasible to probe both real space and k-space structural fluctuations by SµXRD in living biological matter. The non invasive scanning x-ray diffraction method, like Fourier Transform Infrared (FTIR) imaging [42], provides novel perspectives for experimental medicine using synchrotron radiation. Using this approach we have mapped the ultra-structure of a freshly extracted myelinated peripheral nerve, and a nerve that has been cultured for 18 hours with high spatial resolution of 1 µm (Fig. 1C). Whilst somewhat limited by the small sample size (n=1), the probable tissue degradation of myelin we observe following prolonged culture of a myelinated nerve is characterized by the formation of compact and ordered ultrastructural conformations. Further experimentation using the same approach should be carried out to confirm these findings and investigate them further. However, with that caveat, we have described here that the nanostructural detail of the exquisitely ordered structure of myelin begins to alter only a short time following perturbation of the system by removal of a myelinated nerve from an organism.

The biological questions that arise from these observations are several: Firstly whether the fairly subtle changes we have revealed here have functional consequences for myelin complex functionalities [43]. Another question is whether the change we see is actively achieved through an intracellular or extra cellular signaling mechanism or whether, perhaps, it is simply a disturbance in the almost crystalline structure of myelin lamellae secondary to deficiency in blood supply to the nerve. In either of these cases it may be reversible upon restoration of/to a normal physiological support/environment. This may have important consequences for the understanding of acute injury to peripheral nerves and its treatment. These new perspectives can be explored by further investigations into the dynamics, causes and consequences of the compaction and ordered stiffening that we have described. The S*µ*XRD methods we have described here and refinements of them will permit these investigations to be carried out.

**Acknowledgements.** The authors thank Dr. Andre Popov and N.P. acknowledges the Marie Curie Intra European Fellowship for financial support.

**Author contributions:**

N.P., A.R., G.C., A.B, T.A.H. performed the experiments; E. D.-C. provided the XRD station at ESRF; N.P. and T.A.H. prepared the samples; G.C. and A.S.C. followed the data analysis; A.B., N.P., T.A.H., G.C together wrote the paper.
The authors declare no conflict of interest.

**Materials and Methods**

*Methods*. The experimental methods were carried out in "accordance" with the approved guidelines. The source of the synchrotron radiation beam is a 18 mm period in-vacuum undulator, providing 12–13-keV X-rays to crystal optics. The beam is first monochromatized (X-ray energy 12.4 keV) by a liquid nitrogen cooled Si-111 double monochromator; then is focused by a Kirkpatrick–Baez (KB) mirror system to give a flux of $10^{10}$ counts/second on 1x1 µm area. The sample holder hosts the capillary-mounted nerve with the longitudinal axis laying on the vertical x-axis, and allows y-z translation stages. A Fast Readout Low Noise (FReLoN) camera (2048 x 2048 pixel, 50 x 50 µm$^2$ pixel size) is placed at a distance of 820 mm from the sample to collect the 2-D diffraction patterns in transmission. Detector characteristics and specimen-to-film distance were established by recording diffraction patterns from a standard consisting of silver behenate powder ($AgC_{22}H_{43}O_2$), which has a fundamental spacing of $d_{001}$=58.38 Å.

*Materials.* Two adult female frogs (Xenopus laevis; 12 cm length, 180-200g Xenopus express, France) were housed and euthanased at the Grenoble Institute of Neurosciences with the kind cooperation of Dr Andre Popov. The animal experimental protocol was approved by the local committee of the Grenoble Institute of Neurosciences. The frogs were individually transferred in water to a separate room for euthanasia which was carried out using a terminal dose of tricaine (MS222) by immersion followed by decapitation. Terminal anaesthesis was confirmed by the absence of reflexes. The two sciatic nerves were extracted immediately following decapitation.  This was carried out carefully by ligating at two standardized points (proximal and distal to the spine) with silk sutures and cutting outside these points. The proximal point was just distal to where the sciatic nerve forms from spinal roots exiting the column.  The distal point was just proximal to the first branch point.  After dissection, the ligated sciatic nerves were equilibrated in culture medium at pH 7.3 for at least 3 hours at room temperature.  This equilibration was intended to ensure that axons and cells in the nerve had gained osmotic balance with the culture medium. This was less time than previous studies using wide beam x-ray diffraction [19, 21] but because we were interested in examining the acute effects of axonal severing we deemed it a reasonable compromise. The culture medium was a normal Ringer's solution, containing 115 mM NaCl, 2.9 mM KCl, 1.8 mM $CaCl_2$, 5 mM HEPES (4-2-hydroxyethyl-1-piperazinyl-ethanesulfonic). Following equilibration, one of the freshly extracted nerves (henceforth native) was immediately placed in a thin-walled quartz capillary, sealed with wax and mounted on the sample holder for the S$\mu$XRD imaging measurements. The other nerve (henceforth the *early degraded* nerve) was left in the culture medium for 18 hours at room temperature and was finally mounted on the sample holder to perform *µ*XRD in the same day.